\documentclass[12pt]{article}
\usepackage{amsfonts}

\makeatletter   

\def\baselinestretch{2}

\def\setstretch#1{\renewcommand{\baselinestretch}{#1}}

\def\singlespace{\def\baselinestretch{1}\@normalsize}

\def\@setsize#1#2#3#4{\@nomath#1\let\@currsize#1\baselineskip
   #2\baselineskip\baselinestretch\baselineskip
   \setbox\strutbox\hbox{\vrule height.7\baselineskip
      depth.3\baselineskip width\z@}
   \normalbaselineskip\baselineskip#3#4}

\skip\footins 20pt plus4pt minus4pt

\def\@xfloat#1[#2]{\ifhmode \@bsphack\@floatpenalty -\@Mii\else
   \@floatpenalty-\@Miii\fi\def\@captype{#1}\ifinner
      \@parmoderr\@floatpenalty\z@
    \else\@next\@currbox\@freelist{\@tempcnta\csname ftype@#1\endcsname
       \multiply\@tempcnta\@xxxii\advance\@tempcnta\sixt@@n
       \@tfor \@tempa :=#2\do
                        {\if\@tempa h\advance\@tempcnta \@ne\fi
                         \if\@tempa t\advance\@tempcnta \tw@\fi
                         \if\@tempa b\advance\@tempcnta 4\relax\fi
                         \if\@tempa p\advance\@tempcnta 8\relax\fi
         }\global\count\@currbox\@tempcnta}\@fltovf\fi
    \global\setbox\@currbox\vbox\bgroup 
    \def\baselinestretch{1}\@normalsize
    \boxmaxdepth\z@
    \hsize\columnwidth \@parboxrestore}
\long\def\@footnotetext#1{\insert\footins{\def\baselinestretch{1}\footnotesize
    \interlinepenalty\interfootnotelinepenalty 
    \splittopskip\footnotesep
    \splitmaxdepth \dp\strutbox \floatingpenalty \@MM
    \hsize\columnwidth \@parboxrestore
   \edef\@currentlabel{\csname p@footnote\endcsname\@thefnmark}\@makefntext
    {\rule{\z@}{\footnotesep}\ignorespaces
      #1\strut}}}
\makeatother

\newlength{\listindent}
\newlength{\listsep}
   {\begin{list}{ }{%
     \settowidth{\labelwidth}{\textrm{#1}}%
     \setlength{\leftmargin}{\labelwidth }}}%
   {\end{list}}

\newcommand{\eg}{{\it e.g.}} 
\newcommand{\ie}{{\it i.e.}}
\def\Bbb{\mathbb}

\begin{document}

\pagestyle{empty}

\begin{flushright}
  August, 1998\\
  Revised February, 1999
\end{flushright}

\vfill
\begin{center}

 \begin{singlespace}
         {\Large\bfseries A Note on modified Veselov-Novikov Hierarchy}
 \end{singlespace}
 \vspace{.6in}

\normalsize
 {\scshape  Kengo Yamagishi} 
   
\vspace{.2in}
{\itshape
        Akatsuka 3-33-7-202, Itabashi-ku\\ 
        Tokyo 175-0092, JAPAN
}                                  
 \bigskip
 \vfill
 {\bf                     
     ABSTRACT
 }

 \mbox{}

\parbox{5.5in}
 {\hspace{.2in}  Because of its relevance to lower-dimensional
conformal geometry, known as a generalized Weierstrass inducing, the modified
Veselov-Novikov (mVN) hierarchy attracts renewed interest recently.  It has
been shown explicitly in the literature that  an
extrinsic string action \`a la  Polyakov (Willmore functional) is invariant
under deformations associated to the first member of the mVN hierarchy.  In
this note we go one step further and show the explicit invariance of the
functional under deformations associated to all higher members of the
hierarchy.    }

\end{center}

\vfill
\vspace{.5in}
\noindent\hspace{0in} PACS numbers: 11.17, 11.10L, 12.10G

\newpage
\pagestyle{plain}
\parindent.3in

\section{Introduction}

In a last decade we have witnessed a tremendous flow of applications of  
$1(+1$)-dimensional exactly soluble models in 2-d CFT, 2-dimensional
gravity --- both continuum and matrix model approach ---, (super) string
theories, etc.  Specifically, KdV, KP hierarchy and their modified cousins
played important and remarkable roles in various occasions.  We owe this
success mainly to the existence of fascinating mathematical structures
underlying such exactly soluble models, \ie\ an infinite dimensional symmetry,
known as ${\cal W}$-algebras, including Virasoro algebra.

In the case of higher dimensional exactly soluble models (see \eg\
\cite{abl,kono93}), however, even though their importance was
pointed out some time ago \cite{ky}, due to their mathematical complexity, not
much application has been explored until recently. 

Veselov-Novikov (VN) hierarchy \cite{VN} and its modified cousin (mVN)
\cite{bog} are demonstrated as another type of $2(+1)$-dimensional extension of
KdV and mKdV, as compared to the well-known KP-hierarchy.  One interesting
feature of this higher dimensional generalization is that in the mVN case  one
deals with a deformation problem of Dirac operators in 2 dimensions \cite{bog},
rather than that of quadratic differential operators as in ordinary cases.  In
addition, thanks to the contributions of the authors of the recent literature
\cite{kono96,kt,taim} we now know the important relevance of mVN to conformally
Euclidean immersion of 2-surfaces into 3 (or higher) dimensional Euclidean (or
Minkowski) manifold \cite{ken,hoff,eisen}.  There the potential term in
the mVN equation is interpreted as mean curvature of the immersed surface 
(times $\sqrt[4]{g}$), and the first integral of the mVN equation 
(\ie\ 1-st member 
of the hierarchy) is shown \cite{carkon,kt} to be in agreement with Polyakov's
extrinsic string action \cite{poly} (Willmore functional \cite{will}) in
Euclidean signature at classical level.  

Purpose of this note is to show explicitly that the first integral is also
invariant under the deformations associated to the rest of the members of the
hierarchy.  This confirms the statement suggested in the literature
\cite{carkon} obtained from general argument of the soluble system.  
The derived
transformation laws for the potential will also be useful, following similar
methods to  ordinary (intrinsic) string cases, to pin down the algebraic
structure of the infinite symmetries in the extrinsic strings.\footnote[2]{
As a decade-old subject, there are a huge number of works/contributions
to the subject.  The references cited in this Letter may not reflect
all of them.}
(For the current status of the extrinsic strings in connection with QCD, 
see \eg\ \cite{ph} and references therein.)

\section{The mVN and generalized Weierstrass inducing}

We first review generalized Weierstrass inducing
 and see how the mVN hierarchy is involved
there.  The Weierstrass representation is the construction of the conformally
Euclidean minimal surface (\ie\ that with vanishing mean curvature) in 
Euclidean 3-space. (See also \cite{par}.)  The generalized Weierstrass inducing 
considered here is the extension of this construction to non-minimal surfaces.  
To explain this, here we follow the notations of Kenmotsu \cite{ken}.

Let $x_j:\Sigma \rightarrow \Bbb{R}^3\ (j=1,\dots,3)$ be a conformally 
Euclidean immersion of an oriented 2-surface $\Sigma$ (coordinatized locally by
$z,\overline{z}\in \Bbb{C}$) into $\Bbb{R}^3$.  Then following \cite{ken} we 
have:
\begin{eqnarray}
\partial\overline{\partial}x_j &=&
    \frac{\lambda^2}{4}(h_{11}+h_{22})\,e_{3j},\\ 
{\overline{\partial}}^2 x_j &=& \overline{\partial}\lambda\cdot(e_1+i\, e_2)_j
    +\frac{\lambda^2}{4}(h_{11}-h_{22}+2i\,h_{12})\,e_{3j},\\
{\partial}^2 x_j &=& \partial\lambda\cdot(e_1-i\, e_2)_j
    +\frac{\lambda^2}{4}(h_{11}-h_{22}-2i\,h_{12})\,e_{3j}.
\end{eqnarray}
Here $\lambda(>0)$ is related to the conformal factor of the induced metric
\begin{equation}
  ds^2\equiv \sum_{j=1}^3(dx_j)^2=\lambda^2dz\,d\overline{z}\ ,\quad\ {\rm
  or}\quad \ \lambda^2=2\sum_j\left|\frac{\partial x_j}{\partial
  z}\right|^2 =2\sqrt{|g|}\ ,
\end{equation}
and $h_{ij}$ are related to the mean $H$ and Gaussian $K(\sim$ Ricci scalar)
curvatures
\begin{eqnarray}
  H &=& \frac{1}{2} (h_{11}+h_{22})\\
  K &=& H^2 - |\phi|^2\ ,\quad \phi\equiv\frac{1}{2}
  (h_{11}-h_{22})-i\,h_{12}\ , 
\end{eqnarray}
respectively.  The quantities $e_\alpha (\alpha=1,\dots,3)$ are normalized
tangent and normal vectors of the immersed surface $\Sigma$, whose components
are defined as follows: 
\begin{equation}
  e_{1j}=\frac{1}{\lambda}\left(\frac{\, \partial x_j}{\partial z}+
  \frac{\ \partial x_j}{\partial\overline{z}}\right)\, ,\quad 
  e_{2j}=\frac{i}{\lambda}\left(\frac{\, \partial x_j}{\partial z}-
  \frac{\ \partial x_j}{\partial\overline{z}}\right)\, ,\quad
  e_{3i}=\epsilon_{ijk}\,e_{1j}\,e_{2k}\, .
\end{equation}

Next we introduce $\psi_{1,2}$ \cite{taim}
\begin{eqnarray}
  \psi_1 &\equiv& \left[\,\,\,\overline{\partial}(x_2+i\,x_1)\,\right]^{1/2}\ ,
  \label{def1}\\ 
  \psi_2 &\equiv& \left[-\partial(x_2+i\,x_1)\,\right]^{1/2}\ .
  \label{def2} 
\end{eqnarray}
Then by direct calculation, using above formulas, we find that $\psi_i$'s have
to satisfy
\begin{eqnarray}
  \partial\psi_1 &=& \frac{\lambda H}{2}\psi_2 \label{vn1}\\
  \overline{\partial}\psi_2 &=& -\frac{\lambda H}{2}\psi_1 \label{vn2}
\end{eqnarray}
and
\begin{eqnarray}
  \overline{\partial}(\psi_1/\lambda) &=& \frac{\overline{\phi}}{2}\psi_2\\
  {\partial}(\psi_2/\lambda) &=& -\frac{\phi}{2}\psi_1 \ .
\end{eqnarray}
It is also easy to show 
\begin{equation}
  \lambda=|\psi_1|^2+|\psi_2|^2\, .
\end{equation}
It is remarkable that eqs.(\ref{vn1}),(\ref{vn2}) guarantee the integrability
of forms $\Omega_{\pm}, \Omega_3$ defined by
\begin{equation}
  \Omega_+=\overline{\psi}_1^{\ 2}dz -\overline{\psi}_2^{\
  2}d\overline{z}\, ,\quad \Omega_-=\overline{(\Omega_+)}\, ,\quad
  \Omega_3=-(\psi_2\overline{\psi}_1\,dz +
  \psi_1\overline{\psi}_2\,d\overline{z})\, , 
\end{equation}
namely, we have $d\Omega_{\bullet}=0$.  

Now we can consider a converse problem, given a solution to
eqs.(\ref{vn1}),(\ref{vn2}) as a system of differential equations with respect
to $\psi_i$'s.  Since forms $\Omega_{\bullet}$ are all integrable, comparing 
with (\ref{def1}),(\ref{def2}), we observe that
eqs.(\ref{vn1}),(\ref{vn2}) induces an immersion $X_j(z,\overline{z})\
(j=1,\dots,3)$ of conformally Euclidean 2d surface in $\Bbb{R}^3$ via relations 
\begin{equation}
  X_2\mp i\, X_1=\int_\Gamma\Omega_\pm\, ,\quad X_3=\int_\Gamma\Omega_3\, .
\end{equation}
Here $\Gamma$ is an appropriate integration contour ending at
$(z,\overline{z})$.
 The last relation comes from imposed conformally Euclidean property $g_{zz}=
(\partial X_3)^2+(\partial X_1)^2+(\partial X_2)^2=0$, and $g_{\overline{z}\,
\overline{z}}=0$.  The original Weierstrass inducing corresponds to the
special case $H=0$, \ie\ induction for minimal surfaces.  

An important observation made in the recent literature 
\cite{kono96,kt,taim,kono98} is the relevance of eqs.(\ref{vn1}),(\ref{vn2}) to
the $2(+1)$ dimensional exactly soluble mVN system.  The mVN hierarchy is
defined as a deformation problem associated to 2-dimensional Dirac operator
(times some $\gamma$-matrix\footnote[2]{We would rather use following form
(\ref{Dirac}) of the operator for computational simplicity. We could have used
ordinary form for the Dirac operator.  Only $B_n$ in equation (\ref{defo}) get
changed under such redefinition.  In any event essential point is unaffected.
})  ${\cal L}$ with a potential $p=p(z,\overline{z})$:  
\begin{equation}\label{Dirac}  
  {\cal L}=\left(\begin{array}{cr}
    \partial & -p \\
    p & {\overline{\partial}}\end{array}\right)\, ,\quad
  {\cal L}\left(\begin{array}{c}
    \psi_1 \\
    \psi_2 \end{array} 
  \right) = 0\, .
\end{equation}
We note that eqs.(\ref{vn1}),(\ref{vn2}) correspond to taking special potential
$p=\lambda H/2$.  The $n$-th deformation in the hierarchy is defined via
\begin{equation}
  \frac{\delta}{\delta t_n}\left(\begin{array}{c}
    \psi_1 \\
    \psi_2 \end{array} 
  \right)= A_n\left(\begin{array}{c}
    \psi_1 \\
    \psi_2 \end{array} 
  \right)\, ,
\end{equation}
where the deformation operator $A_n$ takes the form
\begin{equation}\label{Xs}
  A_n=\partial^{2n+1}+\sum_{i=0}^{2n-1}X^{(i)}\partial^{i}
  +\overline{\partial}^{\,
  2n+1}+\sum_{i=0}^{2n-1}\tilde{X}^{(i)}\overline{\partial}^{\, i}
\end{equation}
with $X^{(i)}$, $\tilde{X}^{(j)}$ ($i,j=0,\dots,2n-1$) being $2\times2$
matrices.  These matrices are completely determined, together with the  other
matrix-valued differential operator $B_n$ in eq.(\ref{defo}), from the 
compatibility condition:
\begin{equation}\label{defo}
  \left[\ \frac{\delta}{\delta t_n} -A_n\ , \ {\cal L}\ \right]=B_n{\cal L}\, .
\end{equation}
The operator $B_n$ has a similar expression as in the case $A_n$:
\begin{equation}
  B_n=\sum_{i=0}^{2n-1}S^{(i)}\partial^{i}
  +\sum_{i=0}^{2n-1}\tilde{S}^{(i)}\overline{\partial}^{\, i}
\end{equation}
with $2\times2$ matrices $S^{(i)}$, $\tilde{S}^{(j)}$ ($i,j=0,\dots,2n-1$). 
The compatibility condition (\ref{defo}) also gives a deformation equation
for the potential $p$ in the form
$$
  \frac{\delta p}{\delta t_n}= \partial^{2n+1}p +\overline{\partial}^{\, 
  2n+1}p+\cdots\, .
$$

The first case $n=1$ is known to yield modified Veselov-Novikov equation. 
Here we just write down the result.  We will provide more technical details
for higher mVN case in later sections.  
\begin{equation}
  \frac{\delta p}{\delta t_1}= \partial^{3}p + 3\omega\, \partial p+
  \frac{3}{2}p\,\partial\omega+ \overline{\partial}^{\, 3}p+
  3\overline{\omega}\,  \overline{\partial}p+
  \frac{3}{2}p\, \overline{\partial}\overline{\omega}\ ,\quad
  \overline{\partial}\omega\equiv\partial p^2\, .
\end{equation}
It is remarkable that we have a simple first integral of this deformation,
which is obtained from the relation
\begin{equation}
  \frac{\delta p^2}{\delta t_1}=\partial\Bigl(\partial^{2}p^2 - 3(\partial
  p)^2+ 3p^2\omega\Bigr)+ \overline{\partial}\Bigl(\overline{\partial}^{\,2}p^2 
  - 3(\overline{\partial}p)^2+ 3p^2\overline{\omega}\Bigr) .
\end{equation}
Namely, the integral $S$
\begin{equation}\label{will}
  S=2\int p^2\, dz\,d\overline{z}
\end{equation}
does not change its value under the first deformation ${\delta }/\!{\delta
t_1}$ (if $p$ is localized). This conserved quantity has special meaning in
the generalized Weierstrass inducing discussed before.  Substituting
$p=\lambda H/2$, we find that $S$ is nothing but  Polyakov's extrinsic
string action 
$$
  S=\int\sqrt{|g|}\  H^2\, d^2\!x.
$$
(That is known as a Willmore functional in the mathematics literature.)

To sum, Polyakov's extrinsic string action is invariant under the deformation
associated to the (1st) mVN equation.

\section{Second mVN deformation}

We write deformation operators for the 2nd mVN as
\begin{equation}
  \frac{\delta}{\delta t_2}\left(\begin{array}{c}
    \psi_1 \\
    \psi_2 \end{array} 
  \right)= A_2\left(\begin{array}{c}
    \psi_1 \\
    \psi_2 \end{array} 
  \right)\, ,\quad A_2= A_2^{(+)} + A_2^{(-)}\, ,
\end{equation}
where $2\times2$ matrix-valued operators $A_2^{(\pm)}$ are defined as
\begin{equation}
  A_2^{(+)}= \partial^5+V\partial^3 +W\partial^2 + X\partial +Z\, ,\quad
  A_2^{(-)}= \overline{\partial}^{\,5}+\tilde{V} \overline{\partial}^{\,3}
  +\tilde{W} \overline{\partial}^{\,2} + \tilde{X}\overline{\partial}
  +\tilde{Z}\, .
\end{equation}
Then all we have to do is to work out the compatibility condition (\ref{defo})
for $B_2=B_2^{(+)} + B_2^{(-)}$, here represented as
$$
  B_2^{(+)}= Q\partial^3 +R\partial^2 + S\partial +T\, ,\quad
  B_2^{(-)}= \tilde{Q} \overline{\partial}^{\,3}
  +\tilde{R} \overline{\partial}^{\,2} + \tilde{S}\overline{\partial}
  +\tilde{T}\, .
$$
We will perform this independently on $(+)$ and $(-)$ parts of the
compatibility conditions
\begin{equation}\label{comp}
  \left[\ \frac{\delta}{\delta t_2^\pm} -A_2^{(\pm)}\ , \ {\cal L}\
  \right]=B_2^{(\pm)}{\cal L}\, ,\quad\quad  \frac{\delta}{\delta t_2}=
  \frac{\delta}{\delta t_2^+} + \frac{\delta}{\delta t_2^-}\, .
\end{equation}
Some cares must be taken with regards to matrix components $V_{11},W_{11},
X_{11}$,  and $\tilde{V}_{22}, \tilde{W}_{22},\tilde{X}_{22}$.  For instance,
$X_{11}$ and $\tilde{X}_{22}$ give rise to a term in the deformation
\begin{equation}
  \frac{\delta}{\delta t_2}\left(\begin{array}{c}
    \psi_1 \\
    \psi_2 \end{array} 
  \right)\sim \left(\begin{array}{c}
    X_{11}(\partial\psi_1-p\psi_2) \\
    \tilde{X}_{22}(\overline{\partial}\psi_2+p\psi_1) \end{array} 
  \right)\, ,
\end{equation}
which vanishes on-shell, so we set $X_{11}=\tilde{X}_{22}=0$ by hand. 
Similarly we set $V_{11}=W_{11}=\tilde{V}_{22}=\tilde{W}_{22}=0$.  In the 
course of calculation we also get
$Z_{11}=\,\,\stackrel{\circ}{Z}_{11}\,\,=\hbox{\rm const}$, and  
$Z_{22}=\,\,\stackrel{\circ}{Z}_{11}+\cdots$.  This type of deformation 
$$
  \frac{\delta}{\delta t_2}\left(\begin{array}{c}
    \psi_1 \\
    \psi_2 \end{array} 
  \right)\sim \ \stackrel{\circ}{Z}_{11}\!\!\left(\begin{array}{c}
    \psi_1 \\
    \psi_2 \end{array}
  \right)\, 
$$
is merely an overall constant scalar transformation, so we set
$\stackrel{\circ}{Z}_{11}=0$ as well.  Same is true for $\tilde{Z}_{22}$.

After all these done, the deformation matrices are uniquely
determined.  For the operator $A_2^{(+)}$ we obtain
$$
  V=\left(\begin{array}{cc}
    0 & -5\partial p\\
    0 & 5\omega \end{array} 
  \right)\ ,\quad\quad 
  W=\left(\begin{array}{cc}
    0 & -5\partial^2\! p+5p\,\omega\\
    0 & 15\,\partial\omega\!/2 \end{array} 
  \right)\ ,
$$
\begin{equation}\label{result}
  X=\left(\begin{array}{cc}
    0 & \frac{5}{2}(p\partial\omega-2\omega\partial p-2\partial^3\!p)\\
    0 & \frac{5}{2}(2\omega^2+3\,\partial^2\!\omega+2\zeta)
  \end{array} 
  \right)\ ,\quad \overline{\partial}\zeta\equiv\partial(p^2\omega-(\partial
  p)^2)\, ,
\end{equation}
$$
  Z=\left(\begin{array}{cc}
    0 & 5(p\,(\omega^2+\zeta+\partial^2\omega)+\omega\partial^2\!
  p+\frac{1}{2}\partial p\,\partial\omega)\\
    0 & \frac{5}{2}\partial(\omega^2+\zeta+\partial^2\omega)
  \end{array} 
  \right)\ .
$$
For the operator $B_2^{(+)}$ the result is
$$
  Q=\left(\begin{array}{cc}
    0 & -5\partial p\\
    5\,\partial p & 0 \end{array} 
  \right)\ ,\quad\quad 
  R=\left(\begin{array}{cc}
    0 & ((W_{12}))\\
    10\partial^2\! p+5p\,\omega & 0 \end{array} 
  \right)\ ,
$$
\begin{equation}
  S=\left(\begin{array}{cc}
    0 & ((X_{12}))\\
    \frac{5}{2}(3p\partial\omega+6\omega\partial p+4\partial^3\!p) & 0
  \end{array} 
  \right)\ ,\quad 
\end{equation}
$$
  T=\left(\begin{array}{cc}
    0 & ((Z_{12}))\\
    \frac{5}{2}p\,(2\omega^2+2\zeta+3\partial^2\omega)+15\omega\partial^2\!
  p+15\partial p\,\partial\omega + 5\partial^4\!p & 0
  \end{array} 
  \right)\ .
$$
The `12'-components should be taken from those in eq.(\ref{result}).  The 
results for operators $A_2^{(-)}$ and $B_2^{(-)}$ are given by the general rule
\begin{equation}
  A_2^{(-)}=\left(\begin{array}{cc}
    0 & -1\\
    1 & 0 \end{array} 
  \right)\overline{ A_2^{(+)}}\left(\begin{array}{cc}
    0 & 1\\
    -1 & 0 \end{array} 
  \right)\, ,\quad
  B_2^{(-)}=\left(\begin{array}{cc}
    0 & -1\\
    1 & 0 \end{array} 
  \right)\overline{ B_2^{(+)}}\left(\begin{array}{cc}
    0 & 1\\
    -1 & 0 \end{array} 
  \right)\, ,
\end{equation}
where $\overline{ A_2^{(+)}}$ and $\overline{ B_2^{(+)}}$ imply taking complex
conjugate of respective components of the matrix operators.  

Finally the
deformation equation for the potential $p$ is obtained as
\begin{equation}\label{2nd}
  \frac{\delta p}{\delta t^+_2}= \partial^5\!p+ 5\omega\,\partial^3\!p +
  \frac{15}{2}\partial\omega\,\partial^2\!p + \frac{5}{2}\partial p\,(
  2\omega^2+3\,\partial^2\!\omega+2\zeta) + \frac{5}{2}p\,\partial( 
  \omega^2+\zeta+\partial^2\omega)\, ,
\end{equation}
and $\delta p/\delta t^-_2=\overline{\delta p/\delta t^+_2}$, respectively.

At this stage we can check explicit invariance of our first integral
(\ref{will}).  From our deformation equation (\ref{2nd}) we immediately have
\begin{eqnarray}
  \lefteqn{\frac{\delta p^2}{\delta t^+_2}= \partial\Biggl[\partial^4\!p^2-
  5\partial^2(\partial p)^2+ } \nonumber\\ 
  & & +5(\partial^2p)^2 +5\omega\partial^2p^2- 15\omega(\partial p)^2 +
  \frac{5}{2}\partial\omega\,\partial p^2 +
  5 p^2\,(\omega^2+\zeta+\partial^2\omega) 
  \Biggr] ,\label{res2}
\end{eqnarray}
and similarly for $\delta p^2/\delta t^-_2=\overline{\delta p^2/\delta t^+_2}$. 
Thus we find that for localized  $p$ the integral (\ref{will}) remains invariant
under deformations associated to our 2nd member of mVN hierarchy.  

Actually, if the purpose is only to show $\delta p^2/\delta t^\pm_2=$ (total
divergence), the following is more straightforward.  We just inspect `21'- and
`12'-components of every terms from the compatibility conditions (\ref{comp}),
which provide us
\begin{eqnarray}
  5\partial p  +  V_{12}&=&0 \nonumber\\
  10\partial^2p +\partial V_{12}-p\,V_{22}  +  W_{12}&=&0 \nonumber \\
  10\partial^3p +\partial W_{12}-p\,W_{22}  +  X_{12}&=&0  \\
  5\partial^4p +\partial X_{12}-p\,X_{22}  +  Z_{12}&=&0 \nonumber \\
  2\partial^5p+\partial Z_{12} +V_{22}\partial^3p
  +W_{22}\partial^2p+ X_{22}\partial p  &=& 2\delta p/\delta
  t^+_2\, .\nonumber
\end{eqnarray}  
Eventually we end up with a simpler expression
\begin{equation}
  \frac{\delta p^2}{\delta t_2^+}=\partial\left(\partial^4p^2 +V_{12}\partial^3p
  +W_{12}\partial^2p + X_{12}\partial p +Z_{12}p \right).
\end{equation}
This agrees with the previous result (\ref{res2}) after substitutions of the
matrix-components from eqs.(\ref{result}).  The argument for the other
deformation $\delta p^2/ \delta t_2^-$ goes completely in parallel.

From these calculations we can naturally infer that the similar structure
persists in higher mVN deformations, and we can state quite safely that we have
generally 
\begin{equation}
  \frac{\delta p^2}{\delta t_n}=\partial\left(\partial^{2n}p^2+\sum_{i=0}^{2n-1}
  X^{(i)}_{12}\partial^ip\right) +
  \overline{\partial}\left(\overline{\partial}^{\,2n}p^2+\sum_{i=0}^{2n-1}
  \overline{X}^{(i)}_{12}\overline{\partial}^ip\right)
\end{equation}
in our original notation (\ref{Xs}).  On the physics side
this indicates the invariance of the extrinsic action \`a la Polyakov under
all deformations associated to mVN hierarchy.

\section{Discussions}

We have derived 2nd member of the mVN hierarchy, and checked its consistency to
the first integral (one of the so-called Kruskal integrals) derived from the
1st member of the hierarchy.  Even though the result is not unexpected from
the general argument of exactly soluble models of this sort, writing down a
correct form of explicit deformation equation is very important in various
respects.  First of all, we would like to know the complete symmetry structure
of the Polyakov's extrinsic string.  In order to pin down such an infinite
symmetry we have to be aware of the algebraic structure of the Poisson algebra
associated to this system.  In the case of KdV, KP we have successfully
identified their Poisson structures \cite{ds} to be $W_{\infty}$-algebra family
\cite{ky}.  The latter has played an important role in other string analyses.  
Higher Kruskal integrals are also important there.  The derived deformation
equations here are crucial to deduce such Poisson structures.   

Secondly, though the immediate relevance found is to Polyakov's extrinsic
string, we have pointed out several years ago \cite{ky} the importance of 2- (or
higher) dimensional exactly soluble system (that reduces to KdV in
one dimension) in its application to 4-dimensional self-dual gravity.  The (m)VN
is one of the simplest extension of (m)KdV, other than KP.  Clarifying its
symmetry structure through associated Poisson algebra is an important step
towards that goal as well.   We would like to know the possible relevance of mVN
to that problem, before proceeding to more complex Davey-Stewartson hierarchy.  
Though calculations become more involved in higher dimensional exactly soluble
system, we wish to report on our analyses of these issues in future
publications.

\end{document}